\documentclass[aps,preprint,onecolumn,superscriptaddress,nofootinbib,floatfix]{revtex4-1}

\usepackage{graphicx}
\usepackage{dcolumn}
\usepackage{amsmath}
\usepackage{amssymb}
\usepackage{xcolor}
\usepackage{multirow}
\usepackage{listings}
\usepackage{xcolor}
\usepackage{lineno}
\usepackage{extarrows}
\usepackage[colorlinks,citecolor=blue,linkcolor=blue,urlcolor=blue,anchorcolor=blue]{hyperref}
\usepackage{textcomp}
\usepackage[normalem]{ulem}
\usepackage[OT1]{fontenc}
\allowdisplaybreaks
\newcommand\dx{{\rm d}}
\newcommand\p{\partial}
\newcommand\etal{{\it et~al.}}

\newcommand\plb{Phys. Lett. B}

\begin{document}

\title{Quantization of the nonstandard propagating gravitational waves in the cosmological background}
\author{S. X. Tian}
\affiliation{School of Physics and Technology, Wuhan University, 430072 Wuhan, China}
\author{Zong-Hong Zhu}
\affiliation{School of Physics and Technology, Wuhan University, 430072 Wuhan, China}
\affiliation{Department of Astronomy, Beijing Normal University, 100875 Beijing, China}
\date{\today}
\begin{abstract}
  Detections of gravitational wave (GW) stimulate the discussion of how GWs propagate in the expanding Universe. General relativity predicts that GWs are massless and propagate at the speed of light with no extra friction term, which relates to the attenuation of GWs, while some modified gravities may predict a different behavior. The mass and speed terms can be tightly constrained by the GW150914-like and GW170817/GRB 170817A events, respectively. However, the friction term remaining unconstrained. In this paper, we quantize the nonstandard propagating gravitational waves with nonzero friction term in the cosmological background, and study the influence of the friction term on the GW luminosity distance in quantum level, and the initial conditions of perturbations given by inflation. We find the quantum nature of the difference between GW and electromagnetic luminosity distance is graviton particle number non-conservation. For the initial conditions, we obtain an analytical expression of the power spectrum with nonzero friction term for the de Sitter background. In observations, both the GW luminosity distance and primordial GWs can be used to constrain the friction term.
\end{abstract}
\maketitle

\section{Introduction}\label{sec:01}
Recently, gravitational waves (GWs) have been detected \cite{Abbott2016_GW150914,Abbott2017_GW170817,Abbott2018_catalog}, and become a powerful tool to explore cosmology \cite{Schutz1986,Sathyaprakash2009,Abbott2017_H0,Chen2018,Feeney2019,Fishbach2019} and gravity theory \cite{Bettoni2017,Baker2017,Creminelli2017,Ezquiaga2017,Sakstein2017,Akrami2018,Cai2018,Copeland2019,Crisostomi2018,
Gumrukcuoglu2018,Oost2018,Ramos2019}. In general relativity, the linearized Einstein equation shows GWs are massless and propagate at the speed of light with no friction term. However, in some modified gravities that used to explain the cosmological late-time acceleration, the equation of motion of GWs may be different and can be generally written as \cite{Saltas2014,Nishizawa2018,Arai2018,Nishizawa2019}
\begin{align}\label{eq:01}
  \ddot{h}_i+(3+n)H\dot{h}_i+c_T^2\frac{k^2}{a^2}h_i+m_g^2h_i=0,
\end{align}
where $\dot{}\equiv\dx/\dx t$, $H$ is the Hubble parameter, $m_g$ denotes the mass, $c_T$ denotes the speed, and $n$ denotes the friction term. The attenuation of GWs depends on the value of $n$, which is similar to the friction term in the classical mechanics. So, we name $n$ as the friction term as did in \cite{Belgacem2019}. In general relativity, $m_g=0$, $c_T=1$ and $n=0$.

There are many different approaches to constrain $m_g$ (see \cite{Goldhaber2010,deRham2017} for reviews). For example, the gravitational potential of a point source in massive gravity is the Yukawa potential, and thus observations about gravitational bound systems can be used to constrain $m_g$ \cite{Will1998,Finn2002,Zakharov2016}. Ground-based detection of GWs could constrain the graviton mass because nonzero $m_g$ makes the speed of GWs depends on the frequency \cite{Abbott2016_GW150914testGR,Abbott2018_GW170817testGR}. All of these bounds are quite tight ($m_g<10^{-20}{\rm eV}$), and we assume $m_g=0$ hereafter. In addition, primordial GW could also be a possible probe to constrain $m_g$ if detected \cite{Dubovsky2010}.

To constrain the speed of GWs, one can directly compare the arrival time difference with distance between different ground-based detectors \cite{Blas2016,Cornish2017}. However, this bound is very weak. The tightest bound comes from the binary neutron star merger signals, which give $c_T=1\pm\mathcal{O}(10^{-15})$ \cite{Abbott2017_GW-GRB}. We assume $c_T=1$ hereafter. In addition, primordial GW could also be a possible probe to constrain $c_T$ if detected \cite{Amendola2014,Raveri2015,Cai2016}.

An observable effect of the friction term is that the GW luminosity distance $D_L^{({\rm gw})}$ is not equal to the electromagnetic (EM) luminosity distance $D_L^{({\rm em})}$ \cite{Belgacem2018a,Belgacem2018b,Nishizawa2018,Arai2018,Nishizawa2019,Tsujikawa2019}. Any difference between $D_L^{({\rm gw})}$ and $D_L^{({\rm em})}$ indicates new physics. Previously, people proved this with the classical field theory. Here, we ask what is the quantum nature of $D_L^{({\rm gw})}\neq D_L^{({\rm em})}$. Note that, there may be other reasons to be responsible for $D_L^{({\rm gw})}\neq D_L^{({\rm em})}$, e.g., high spacetime dimensions \cite{Pardo2018}, time-varying Planck mass \cite{Amendola2018,Lagos2019}, quantum gravity dimensional flow \cite{Calcagni2019}, viscous Universe \cite{Lu2018}, modified redshift relation \cite{Bassett2013,Wojtak2016,Wojtak2017,Tian2017} and so on. In this paper, we focus on the friction term caused by modified gravities, i.e., we assume the dimension of spacetime is $3+1$ and all fundamental constants are same as in the classical quantum field theory. Especially, we assume $n$ is constant.

This paper is organized as follows: Section \ref{sec:02} canonically quantizes the nonstandard propagating field with constant $n$. Sections \ref{sec:03} and \ref{sec:05} analyze the influence of the friction term on $D_L^{({\rm gw})}$ in quantum level and the power spectrum of initial perturbations given by the inflationary theory, respectively. Our conclusions will be presented in Sec. \ref{sec:06}. Conventions: $i=\sqrt{-1}$ and $c=\hbar=8\pi G=1$.

\section{Canonical quantization}\label{sec:02}
In this section, we canonically quantize the nonstandard propagating field in order to explore the effects of the friction term on the quantum nature of the field. Our quantization procedure follows \cite{Lancaster2014,Parker1977}. We assume the Universe is described by the flat Friedmann-Lema\^{i}tre-Robertson-Walker (FLRW) metric
\begin{equation}
  \dx s^2=\dx t^2-a^2(\dx x^2+\dx y^2+\dx z^2).
\end{equation}
Until now, we know the equation of motion of the field, i.e., Eq. (\ref{eq:01}) with $m_g=0$, $c_T=1$ and $n\neq0$. However, it is not enough to quantize the field. We still need the Lagrangian density to define the conjugate momentum. In order to obtain the desired equation of motion, we find the Lagrangian density can be written as
\begin{equation}\label{eq:03}
  \mathcal{L}=\sqrt{-g}\frac{a^n}{a_0^n}\cdot\frac{1}{2}g^{\mu\nu}\p_\mu\phi\p_\nu\phi,
\end{equation}
where $a_0$ and $n$ are constant, and $\phi$ denotes a real scalar field. Intuitively, $\mathcal{L}/\sqrt{-g}$ defined in Eq. (\ref{eq:03}) is not a scalar as the scale factor is expressed in the comoving time coordinate. However, it is not hard to rewrite Eq. (\ref{eq:03}) in the manifestly covariant form. Fox example, we can replace the coefficient $a^n/a_0^n$ with $\rho_0/\rho$, where $\rho_0$ is constant and $\rho$ is the energy density of one certain type of perfect fluid with the equation of state $w=p/\rho=-1+n/3$, which gives $\rho\propto a^{-3(w+1)}=a^{-n}$ for the FLRW metric. Note that $\rho$ is a scalar and independent of $\phi$. As we assumed the Universe is described by the flat FLRW metric, we can substitute $\rho\propto a^{-n}$ into the Lagrangian density before using the variational method to derive the $\phi$-field equations. Thus, it is reasonable to write down Eq. (\ref{eq:03}) directly. The discussion here indicates that the friction term is related to the non-minimal coupling of two different fields. In this paper, we do not have to distinguish between the symbols of $h_i$ and $\phi$ because $h_i$ is also a real field and the propagations of GWs with different polarizations are independent of each other. $a_0$ is used to tuning the dimension of $\mathcal{L}$. Variation of the action ($S\equiv\int\dx^4x\mathcal{L}$) with respect to the field gives
\begin{equation}\label{eq:04}
  \frac{\p^2\phi}{\p t^2}+(3+n)H\frac{\p\phi}{\p t}-\frac{1}{a^2}\nabla^2\phi=0.
\end{equation}
Taking a time coordinate transformation
\begin{equation}\label{eq:05}
  \tau=\int^t\frac{a_0^n}{a^{n+3}}\dx t,
\end{equation}
we obtain
\begin{equation}\label{eq:06}
  a_0^{2n}\frac{\p^2\phi}{\p\tau^2}-a^{2n+4}\nabla^2\phi=0.
\end{equation}
Eq. (\ref{eq:05}) applies to Sec. \ref{sec:03}, and we take a new time coordinate transformation in Sec. \ref{sec:05}.

Now, we start the canonical quantization. The mode expansion of the field operator can be written as
\begin{equation}\label{eq:07}
  \hat{\phi}=\frac{1}{(2\pi)^{3/2}}\int\dx^3\mathbf{k}\left[\hat{a}_{\mathbf{k}}e^{i\mathbf{k}\cdot\mathbf{x}}\psi_k(\tau)
  +{\rm H.C.}\right],
\end{equation}
where H.C. denotes the Hermitian conjugate. The subscript of $\psi_k$ is $k$ (not $\mathbf{k}$) means the energy depends only on the modulus (not the direction) of the wave vector. Eq. (\ref{eq:06}) gives
\begin{equation}\label{eq:08}
  a_0^{2n}\frac{\dx^2\psi_k}{\dx\tau^2}+a^{2n+4}k^2\psi_k=0.
\end{equation}
The conjugate momentum operator is
\begin{align}\label{eq:09}
  \hat{\pi}&\equiv\frac{\p\mathcal{L}}{\p(\p_0\phi)}=\frac{a^{n+3}}{a_0^n}\p_0\hat{\phi},\nonumber\\
  &=\frac{1}{(2\pi)^{3/2}}\int\dx^3\mathbf{k}\left[\hat{a}_{\mathbf{k}}e^{i\mathbf{k}\cdot\mathbf{x}}\frac{\p\psi_k}{\p\tau}
  +{\rm H.C.}\right].
\end{align}
The canonical commutation relations are
\begin{gather}
  [\hat{\phi}(\mathbf{x},t),\hat{\phi}(\mathbf{x}',t)]=0,\quad
  [\hat{\pi}(\mathbf{x},t),\hat{\pi}(\mathbf{x}',t)]=0,\nonumber\\
  [\hat{\phi}(\mathbf{x},t),\hat{\pi}(\mathbf{x}',t)]=i\delta^{(3)}(\mathbf{x}-\mathbf{x}'),\label{eq:10}
\end{gather}
which correspond to\footnote{One can substitute Eqs. (\ref{eq:07}) and (\ref{eq:09}) into Eq. (\ref{eq:10}) to obtain Eq. (\ref{eq:11}). However, this is not a one-to-one correspondence. An adjustable constant should appear in Eq. (\ref{eq:11}). We set it to be $1$ as the final physical results are independent of this setting. Note that, this adjustable constant should also exist in the quantization of the standard propagating fields under the Minkowski background \cite{Lancaster2014} and the FLRW background \cite{Parker1977}.}
\begin{subequations}\label{eq:11}
 \begin{gather}
  [\hat{a}_{\mathbf{k}},\hat{a}_{\mathbf{k}'}]=0,\quad
  [\hat{a}_{\mathbf{k}},\hat{a}_{\mathbf{k}'}^\dag]=\delta^{(3)}(\mathbf{k}-\mathbf{k}'),\label{eq:11a}\\
  \psi_k\frac{\dx\psi_{k}^*}{\dx\tau}-\psi_k^*\frac{\dx\psi_{k}}{\dx\tau}=i,\label{eq:11b}
 \end{gather}
\end{subequations}
where $\delta^{(3)}$ means the three dimensional Dirac delta function, ${}^\dag$ means Hermitian conjugate, and ${}^*$ means complex conjugate.

In order to define the vacuum, we can assume the expansion of the Universe is slow enough in a certain period of time (equivalently, the modulus of the wave vector that we are considering is large enough, i.e., the wavelength is short). This assumption allows us to define some concepts in the curved spacetime with the help of the classical quantum field theory (Minkowski metric), and to omit $\dot{a}$-like terms in the following calculations. Based on the classical quantum field theory, we know if $\psi_k$ only contain the negative frequency component, then $\hat{a}_{\mathbf{k}}$ is the annihilation operator, i.e., $\hat{a}_{\mathbf{k}}|0\rangle=0$, which also defines the vacuum. For the pure negative frequency component, Eqs. (\ref{eq:08}) and (\ref{eq:11b}) give
\begin{align}
  \psi_k&=\frac{1}{\left(2\frac{a^{n+3}}{a_0^n}\omega_k\right)^{1/2}}\exp(-i\frac{a^{n+3}}{a_0^n}\omega_k\tau),\nonumber\\
    &\xlongequal{\textrm{phase shift}}
    \frac{1}{\left(2\frac{a^{n+3}}{a_0^n}\omega_k\right)^{1/2}}\exp(-i\omega_k t),\label{eq:12}
\end{align}
where the temporal angular frequency $\omega_k=k/a$, and the second line used Eq. (\ref{eq:05}) and omitted a possible phase difference as it is unnecessary at here.

As did in \cite{Parker1977} (see Eq. (3.21) in \cite{Parker1977}), we still need to find the expression of the field operator in the physical coordinates. The rescaled physical coordinate $\mathbf{y}=a\mathbf{x}$, the physical momentum $\mathbf{p}=\mathbf{k}/a$, and the physical frequency $\omega_p=p\,(=\omega_k)$. As in the classical quantum field theory, we require the physical creation and annihilation operators satisfy
\begin{align}\label{eq:13}
  [\hat{a}_\mathbf{p},\hat{a}_{\mathbf{p}'}]=0,\quad
  [\hat{a}_\mathbf{p},\hat{a}_{\mathbf{p}'}^\dag]=\delta^{(3)}(\mathbf{p}-\mathbf{p}').
\end{align}
$\hat{a}_{\mathbf{p}}^\dag|0\rangle$ means a monochromatic wave with physical momentum $\mathbf{p}$. Comparing Eq. (\ref{eq:13}) with Eq. (\ref{eq:11a}), and recalling the scaling property of the Dirac delta function, we obtain
\begin{equation}\label{eq:14}
  \hat{a}_\mathbf{p}=a^{3/2}\hat{a}_\mathbf{k}.
\end{equation}
This is same as the result in \cite{Parker1977}. Note that, in principle, $\hat{a}_\mathbf{p}$ and $\hat{a}_\mathbf{k}$ are two different operators. Here we use the same symbol $\hat{a}$, and one can distinguish them from the subscripts. Substituting Eqs. (\ref{eq:12}) and (\ref{eq:14}) into Eq. (\ref{eq:07}), we obtain
\begin{align}\label{eq:15}
  \hat{\phi}&=\frac{1}{(2\pi)^{3/2}}\frac{a_0^{n/2}}{a^{n/2}}
  \int\frac{\dx^3\mathbf{p}}{\left(2\omega_p\right)^{1/2}}\left[\hat{a}_\mathbf{p}e^{-ipy}+{\rm H.C.}\right],
\end{align}
where $py=\omega_p t-\mathbf{p}\cdot\mathbf{y}$. Except for one coefficient $a_0^{n/2}/a^{n/2}$, Eq. (\ref{eq:15}) is identical to the results in the standard propagating case \cite{Parker1977}. As in the classical quantum field theory, the particle number density operator $\hat{\rho}(\mathbf{x})=\hat{\phi}^\dag(\mathbf{x})\hat{\phi}(\mathbf{x})$ \cite{Lancaster2014}. Thus, the coefficient $a_0^{n/2}/a^{n/2}$ means the total particle number $N\propto a^{-n}$ with the expansion of the Universe. Note that, this conclusion only holds for the case of slow expansion (or equivalently short wavelength), because Eq. (\ref{eq:15}) only holds for this case.

In order to complete the quantum field theory in curved spacetime for the nonstandard propagating fields, here we discuss the influence of the friction term on the cosmological particle creation. For the standard propagating fields, \cite{Schrodinger1939,Parker1968,Parker1969,Parker1971} first pointed out that the expansion of the Universe could create particles from the vacuum. The analog of this phenomenon in the black hole case is the famous Hawking radiation \cite{Hawking1974,Hawking1975}. For the nonstandard propagating case, as in \cite{Parker1977}, we assume
\begin{equation}\label{eq:17}
  a(\tau)=\left\{
  \begin{array}{ll}
    a_1 & \textrm{when }\tau\rightarrow-\infty,\\
    a_2 & \textrm{when }\tau\rightarrow+\infty,
  \end{array}\right.
\end{equation}
where $a_1$ and $a_2$ are positive constant. The expansion of the Universe is specified as $a=a(\tau)$. The general mode expansion of the field operator is Eq. (\ref{eq:07}), and $\psi_k$ satisfies Eq. (\ref{eq:08}). If we want $\hat{a}_{\mathbf{k}}$ to be an annihilation operator at $\tau=-\infty$, then we require [see Eq. (\ref{eq:12})]
\begin{equation}\label{eq:18}
  \psi_k|_{\tau\rightarrow-\infty}=
  \frac{1}{\left(2\frac{a_1^{n+3}}{a_0^n}\omega_{k,1}\right)^{1/2}}\exp(-i\frac{a_1^{n+3}}{a_0^n}\omega_{k,1}\tau),
\end{equation}
where $\omega_{k,1}=k/a_1$. This can be regarded as the boundary condition in our calculation. The analysis are performed in the Heisenberg picture, where the state is fixed and the operator is evolving with time. We denote $|0_i\rangle$ as the vacuum state at $\tau=-\infty$, and this state is fixed with time increasing. $\hat{a}_{\mathbf{k}}$ is the annihilation operator at $\tau=-\infty$ means $\hat{a}_{\mathbf{k}}|0_i\rangle=0$, and this equality holds for any time. Note that, the operator is evolving in Heisenberg picture does not means $\hat{a}_{\mathbf{k}}$ depends on time, and all the time dependence are including in $\psi_k$. When $\tau\rightarrow+\infty$, the mode expansion can be written as
\begin{align}
  \hat{\phi}|_{\tau\rightarrow+\infty}&=
  \frac{1}{(2\pi)^{3/2}}\int\dx^3\mathbf{k}\frac{1}{\left(2\frac{a_2^{n+3}}{a_0^n}\omega_{k,2}\right)^{1/2}}\times\nonumber\\
  &\qquad\qquad\quad\left[\hat{A}_{\mathbf{k}}e^{i\mathbf{k}\cdot\mathbf{x}}e^{-i\omega_{k,2}t}+{\rm H.C.}\right],\label{eq:19}
\end{align}
where $\omega_{k,2}=k/a_2$, and $\hat{A}_{\mathbf{k}}$ is the annihilation operator at $\tau=+\infty$. Then the number density of the cosmological created particles at $\tau=+\infty$ is
\begin{equation}\label{eq:20}
  n_{\mathbf{k}}=\langle{}_i0|\hat{A}_{\mathbf{k}}^\dag\hat{A}_{\mathbf{k}}|0_i\rangle.
\end{equation}
To calculate $n_{\mathbf{k}}$, we need to find the relation between $\hat{A}_{\mathbf{k}}$ and $\hat{a}_{\mathbf{k}}$. When $\tau\rightarrow+\infty$, the general solution of $\psi_k$ is
\begin{align}
  \psi_k&=\frac{1}{\left(2\frac{a_2^{n+3}}{a_0^n}\omega_{k,2}\right)^{1/2}}\left[\alpha_k\exp(-i\frac{a_2^{n+3}}{a_0^n}\omega_{k,2}\tau)\right.\nonumber\\
  &\qquad\qquad\qquad\left.+\beta_k\exp(i\frac{a_2^{n+3}}{a_0^n}\omega_{k,2}\tau)\right].\label{eq:21}
\end{align}
The expressions of $\alpha_k$ and $\beta_k$ can be obtained from solving Eq. (\ref{eq:08}) with the boundary condition Eq. (\ref{eq:18}). Eq. (\ref{eq:11b}) gives $|\alpha_k|^2-|\beta_k|^2=1$. Substituting Eq. (\ref{eq:21}) into Eq. (\ref{eq:07}), and regrouping the negative and positive frequency components, we obtain
\begin{equation}\label{eq:22}
  \hat{A}_{\mathbf{k}}=\alpha_k\hat{a}_{\mathbf{k}}+\beta_k^*\hat{a}^\dag_{-\mathbf{k}},
\end{equation}
which is identical to Eq. (3.25) in \cite{Parker1977}. Substituting Eq. (\ref{eq:22}) into Eq. (\ref{eq:20}), we obtain $n_{\mathbf{k}}=|\beta_k|^2$.

\section{GW luminosity distance: Particle approach}\label{sec:03}
GW observations provide a direct way to determine the GW luminosity distance $D_L^{({\rm gw})}$, and thus to constrain the cosmological parameters \cite{Schutz1986,Sathyaprakash2009}. GW170817 is the first example of measuring the Hubble constant in this way \cite{Abbott2017_H0,Fishbach2019}, and high precision measurements are possible in the near future \cite{Chen2018,Feeney2019}. In generally relativity, $D_L^{({\rm gw})}$ is equal to the EM luminosity distance $D_L^{({\rm em})}$. However, in some modified gravities, the friction term appears in the GW propagation equations, which results in $D_L^{({\rm gw})}\neq D_L^{({\rm em})}$ \cite{Belgacem2018a,Belgacem2018b,Nishizawa2018,Arai2018,Nishizawa2019,Tsujikawa2019}. Previous calculations mainly focused on how the friction term affects the GW amplitude, i.e., calculated $D_L^{({\rm gw})}$ in the classical field approach. In this section, we revisit the GW luminosity distance in particle approach.

The wavelength of GW170817-like signals is much smaller than the cosmological scale, which means the results obtained in the previous section under slow expansion assumption apply here. So, the quantum nature of $D_L^{({\rm gw})}\neq D_L^{({\rm em})}$ is that the number of graviton particles is not conserved, and the total number $N^{({\rm g})}\propto a^{-n}$. Following the work of \cite{Weinberg2008,Tian2017} that calculated $D_L^{({\rm em})}$ in particle approach, we obtain
\begin{equation}\label{eq:16}
  D_L^{({\rm gw})}=\sqrt{\frac{N^{({\rm g})}(z)}{N^{({\rm g})}({\rm today})}}D_L^{({\rm em})}
  =(1+z)^{n/2}D_L^{({\rm em})},
\end{equation}
where we used the classical redshift relation $1+z=a_{\rm today}/a$. Eq. (\ref{eq:16}) shows if $n>0$, then $D_L^{({\rm gw})}>D_L^{({\rm em})}$, which means the source looks dimmer in the GW channel than in the EM channel. This is consistent with the results obtained in the classical field approach, where a positive $n$ attenuates the GWs faster and makes the source dimmer. Quantitatively, Eq. (\ref{eq:16}) is consistent with the results that presented in \cite{Amendola2018,Nishizawa2018}. As discussed in \cite{Amendola2018}, LIGO cannot provide tight constraints on $n$, but LISA can constrain $n$ with an error $\sigma_n\approx0.13$ in 5 years' operation.

\section{Inflation: Initial conditions}\label{sec:05}
Inflation is proposed to solve the horizon and flatness problems that exist in the classical Big Bang cosmology \cite{Guth1981}. An inborn ability of the inflationary theory is to naturally present the nearly-scale-invariant spectrum of perturbations, which is the initial conditions of cosmic inhomogeneities \cite{Maggiore2000,Dodelson2003,Bassett2006}. In this section, we explore the influence of the friction term on initial spectrum of perturbations. Especially, we focus on the tensor perturbations.

The inflation background is assumed to be the de Sitter Universe
\begin{equation}
  a(t)=a_1\exp(H_i[t-t_1]),
\end{equation}
where $t_1$ is the start time of inflation, $a_1$ is the value of the scale factor at $t=t_1$, and $H_i$ is the Hubble parameter. We denote the end time of inflation as $t_2$, and $a_2=a(t_2)$. The size of causally connected region during inflation is
\begin{align}
  \Delta x_{12}
  =\int_{t_1}^{t_2}\frac{\dx t}{a}
  =\frac{1}{H_i}(\frac{1}{a_1}-\frac{1}{a_2})
  \approx\frac{1}{a_1H_i}.
\end{align}
We denote the size of the EM observable Universe at today as $\Delta x_{34}$. In order to solve the horizon and flatness problems, we require $a_2/a_1\gtrsim\exp(60)$ [or $a_2/a_1\gg\exp(60)$ for a more satisfactory solution to the horizon problem], which ensures $\Delta x_{12}\gtrsim\Delta x_{34}$ [or $\Delta x_{12}\gg\Delta x_{34}$]. Recalling the factor $e^{i\mathbf{k}\cdot\mathbf{x}}$ appears in the Fourier transformation and the anisotropy size of the cosmic microwave background radiations, we know $k\Delta x_{34}\approx100$ gives the characteristic value of the wave number that we are observing. Thus, the following approximations
\begin{equation}\label{eq:29}
  k\gg a_1H_i\quad\textrm{and}\quad\frac{k}{a_1H_i}\ll e^{H_i(t_2-t_1)}
\end{equation}
are reasonable in the perturbation analysis at here.

Quantum fluctuation is the source of initial perturbations. The equation of motion is Eq. (\ref{eq:04}), and the mode expansion of the field operator is still Eq. (\ref{eq:07}). Here, we take a new time coordinate transformation
\begin{equation}
  \tau\equiv\int_{t_1}^t\frac{\dx t}{a}=\frac{1}{a_1H_i}\left[1-e^{-H_i(t-t_1)}\right],
\end{equation}
and then
\begin{equation}
  a=\frac{1}{1-a_1H_i\tau}.
\end{equation}
After taking the above time coordinate transformation and a function transformation ($\psi_k=\tilde{\psi}_k/a^{1+n/2}$), the field equation can be written as
\begin{equation}\label{eq:32}
  \tilde{\psi}_k''+\left(k^2-\frac{2n+n^2}{4}\mathcal{H}^2-\frac{2+n}{2}\cdot\frac{a''}{a}\right)\tilde{\psi}_k=0,
\end{equation}
where $'\equiv\dx/\dx\tau$ and $\mathcal{H}\equiv a'/a$. For the de Sitter background, we know Eq. (\ref{eq:32}) is the Whittaker equation (see Appendix \ref{sec:AppA}), and the general solution is
\begin{equation}\label{eq:33}
  \tilde{\psi}_k=C_1M_{0,(3+n)/2}(2ik\tilde{\tau})+C_2W_{0,(3+n)/2}(2ik\tilde{\tau}),
\end{equation}
where $C_1$ and $C_2$ are constant, the definition of the Whittaker $M/W$ function can be found in Appendix \ref{sec:AppA}, and $\tilde{\tau}=\tau-1/(a_1H_i)$. As all the $k$ we are considering can be easily causally connected during the early stage of inflation\footnote{Mathematically, the early stage means $\tau\ll1/(a_1H_i)$, and the causally connected means $k\tau\gtrsim1$.}, the boundary condition at here should be (see Eq. (\ref{eq:12}) for the coefficient)
\begin{equation}\label{eq:34}
  \psi_k|_{\tau\rightarrow0}=\frac{1}{\left(2\frac{a_1^{n+3}}{a_0^n}\frac{k}{a_1}\right)^{1/2}}\exp(-ik\tau).
\end{equation}
Combined Eq. (\ref{eq:34}) and the results in Appendix \ref{sec:AppB}, we obtain
\begin{equation}
  C_1=0,\quad C_2=\frac{a_0^{n/2}}{\sqrt{2k}}e^{-\frac{ik}{a_1H_i}},
\end{equation}
and then
\begin{align}
  &\psi_k=\frac{a_0^{n/2}}{\sqrt{2k}}e^{-\frac{ik}{a_1H_i}}\cdot\frac{1}{a^{1+n/2}}\cdot
    W_{0,(3+n)/2}(2ik\tilde{\tau}),\nonumber\\
  &\xlongequal{t\rightarrow t_2} a_0^{n/2}\cdot\frac{e^{-\frac{ik}{a_1H_i}}}{\sqrt{2k}}\cdot
    \frac{\Gamma(3+n)}{\Gamma(2+n/2)}\left(\frac{H_i}{-2ik}\right)^{1+n/2},\label{eq:36}
\end{align}
where $\Gamma(z)$ is the gamma function, and the last line used Eq. (\ref{eq:29}), which means $\lim_{t\rightarrow t_2}|k\tilde{\tau}|\ll1$, and Eq. (\ref{eq:45}) in Appendix \ref{sec:AppC}. Thus the power spectrum
\begin{align}\label{eq:37}
  P_\phi(k)=|\psi_k|^2
  =\frac{H_i^2}{2k^3}\frac{\Gamma^2(3+n)}{2^{2+n}\Gamma^2(2+n/2)}\left(\frac{a_0H_i}{k}\right)^n,
\end{align}
which shows the friction term would influence the spectrum index. The above calculations apply to the tensor perturbations, and the results provide a possibility to constrain $n$ through the detection of primordial GWs. For $n=0$, Eq. (\ref{eq:37}) recovers the classical result \cite{Dodelson2003}. For $n\neq0$, Eq. (\ref{eq:37}) is consistent with the result obtain in \cite{Lin2016}, although the calculation details are slightly different.

\section{Conclusions}\label{sec:06}
In this paper, we quantize the nonstandard propagating gravitational waves with nonzero friction term. The widely discussed inequality $D_L^{({\rm gw})}\neq D_L^{({\rm em})}$ induced by the friction term inspire our work. After quantizing the fields under slow expansion approximation (equivalently short wavelength approximation), we point out the quantum nature of $D_L^{({\rm gw})}\neq D_L^{({\rm em})}$ is the particle number non-conservation of graviton. For the inflation with de Sitter background, we obtain the full solution of the perturbation equation, and quantitatively calculate the influence of the friction term on the spectrum index.

\section*{Acknowledgements}
This work was supported by the National Natural Science Foundation of China under Grants No. 11633001 and the Strategic Priority Research Program of the Chinese Academy of Sciences, Grant No. XDB23000000.

\appendix
\renewcommand\thesubsection{\Alph{subsection}}
\section*{Appendix: Whittaker function}
\subsection{Definition of the Whittaker equation and functions}\label{sec:AppA}
As in \cite{Gradshteyn2007}, the Whittaker equation reads
\begin{equation}
  \frac{\dx^2W}{\dx z^2}+\left(-\frac{1}{4}+\frac{\lambda}{z}+\frac{1/4-\mu^2}{z^2}\right)W=0,
\end{equation}
and two linear independent solutions are
\begin{align}
  M_{\lambda,\mu}(z)&=z^{\mu+\frac{1}{2}}e^{-z/2}{}_1F_1(\mu-\lambda+\frac{1}{2};2\mu+1;z),\label{eq:39}\\
  W_{\lambda,\mu}(z)&=\frac{\Gamma(-2\mu)}{\Gamma(\frac{1}{2}-\mu-\lambda)}M_{\lambda,\mu}(z)\nonumber\\
  &\qquad\quad+\frac{\Gamma(2\mu)}{\Gamma(\frac{1}{2}+\mu-\lambda)}M_{\lambda,-\mu}(z),\label{eq:40}
\end{align}
where the confluent hypergeometric function
\begin{equation}\label{eq:41}
  {}_1F_1(\alpha;\gamma;z)=1+\frac{\alpha}{\gamma}\frac{z}{1!}+\frac{\alpha(\alpha+1)}{\gamma(\gamma+1)}\frac{z^2}{2!}+\cdots.
\end{equation}
In the following calculations, we will use the software Maple, in which ${\rm WhittakerM}(\lambda,\mu,z)\equiv M_{\lambda,\mu}(z)$ and ${\rm WhittakerW}(\lambda,\mu,z)\equiv W_{\lambda,\mu}(z)$.

\subsection{Asymptotic behavior of the Whittaker solutions at the beginning of inflation}\label{sec:AppB}
In Sec. \ref{sec:05}, we obtain the solution Eq. (\ref{eq:33}). Here we study the behavior of this solution when $\tau\rightarrow0$. In addition, we assume $|n|\lesssim\mathcal{O}(1)$, i.e., the value of $n$ is not far away from zero. We are unable to directly obtain the final results for the general $n$. So, we start from some special $n$. With the help of Maple, we obtain
\begin{subequations}
\begin{align}
  M_{0,\frac{1}{2}}(2ik\tilde{\tau})&=e^{ik\tilde{\tau}}-e^{-ik\tilde{\tau}},\\
  M_{0,\frac{3}{2}}(2ik\tilde{\tau})
    &=6(e^{ik\tilde{\tau}}+e^{-ik\tilde{\tau}})+\frac{6ia_1H_i(e^{ik\tilde{\tau}}-e^{-ik\tilde{\tau}})}{k(a_1H_i\tau-1)}\nonumber\\
    &\approx6(e^{ik\tilde{\tau}}+e^{-ik\tilde{\tau}}),\\
  M_{0,\frac{5}{2}}(2ik\tilde{\tau})
    &=\left[60-\frac{180a_1^2H_i^2}{k^2(a_1H_i\tau-1)^2}\right](e^{ik\tilde{\tau}}-e^{-ik\tilde{\tau}})\nonumber\\
    &\qquad\quad+\frac{180ia_1H_i}{k(a_1H_i\tau-1)}(e^{ik\tilde{\tau}}+e^{-ik\tilde{\tau}})\nonumber\\
    &\approx60(e^{ik\tilde{\tau}}-e^{-ik\tilde{\tau}}),\\
  M_{0,\frac{7}{2}}(2ik\tilde{\tau})&\approx840(e^{ik\tilde{\tau}}+e^{-ik\tilde{\tau}}),
\end{align}
\end{subequations}
where the approximate equality used $k\gg a_1H_i$, $\tau\ll1/(a_1H_i)$ and $k\tau\gtrsim1$ as discussed in Sec. \ref{sec:05}. The above results indicate the positive frequency component appears in the Whittaker M function (and also $\psi_k$), and we do not want this term as required by the boundary condition Eq. (\ref{eq:34}). For the Whittaker W function, Maple gives
\begin{subequations}\label{eq:43}
\begin{align}
  W_{0,\frac{1}{2}}(2ik\tilde{\tau})&=e^{-ik\tilde{\tau}},\\
  W_{0,\frac{3}{2}}(2ik\tilde{\tau})&=\left[1-\frac{ia_1H_i}{k(a_1H_i\tau-1)}\right]e^{-ik\tilde{\tau}},\nonumber\\
    &\approx e^{-ik\tilde{\tau}},\\
  W_{0,\frac{5}{2}}(2ik\tilde{\tau})&=\left[1-\frac{3ia_1^2H_i^2k\tau+3a_1^2H_i^2-3ia_1H_ik}{k^2(a_1^2H_i^2\tau^2-2a_1H_i\tau+1)}\right]e^{-ik\tilde{\tau}},\nonumber\\
    &\approx e^{-ik\tilde{\tau}},\\
  W_{0,\frac{7}{2}}(2ik\tilde{\tau})&\approx e^{-ik\tilde{\tau}}.
\end{align}
\end{subequations}
The above results indicate
\begin{equation}\label{eq:44}
  \lim_{\tau\rightarrow0}W_{0,(3+n)/2}(2ik\tilde{\tau})=e^{-ik\tilde{\tau}}
\end{equation}
holds for $n>-3$. However, the examples presented in Eq. (\ref{eq:43}) are just for the even $n$. Here we numerically verify Eq. (\ref{eq:44}) for the decimal $n$. We denote $k=n_1a_1H_i$ and $\tau=1/(n_2a_1H_i)$, and the approximations used before correspond to $n_1\gg1$ and $n_2\gg1$. We define $\Delta(n,n_1,n_2)=\left|W_{0,(3+n)/2}(2ik\tilde{\tau})-e^{-ik\tilde{\tau}}\right|$. Using Maple, one can easily verify $\lim_{n_1,n_2\rightarrow+\infty}\Delta(n,n_1,n_2)=0$ for $n=\pm0.1,0.2,0.3,\cdots$. This results indicate Eq. (\ref{eq:44}) holds for the decimal $n$. Our proof for Eq. (\ref{eq:44}) is not mathematically rigorous. But the evidences that provided to support Eq. (\ref{eq:44}) are quite convincing, and we think the result is mathematically correct.

\subsection{Taylor expansion of $W_{0,\mu}(z)$ around $z=0$}\label{sec:AppC}
When $t\rightarrow t_2$, the variable of the Whittaker W function in Eq. (\ref{eq:36}) is close to zero. So, we need the Taylor expansion of $W_{0,\mu}(z)$ around $z=0$. Using Eqs. (\ref{eq:39}--\ref{eq:41}), we obtain
\begin{align}
  \lim_{z\rightarrow0}W_{0,\mu}(z)&=\frac{\Gamma(-2\mu)}{\Gamma(\frac{1}{2}-\mu)}z^{\mu+\frac{1}{2}}+\frac{\Gamma(2\mu)}{\Gamma(\frac{1}{2}+\mu)}z^{-\mu+\frac{1}{2}}\nonumber\\
  &=\frac{\Gamma(2\mu)}{\Gamma(\frac{1}{2}+\mu)}z^{-\mu+\frac{1}{2}},\label{eq:45}
\end{align}
where the last equality used $\mu>0$, which corresponds to $n>-3$ in Sec. \ref{sec:05}.

%

\end{document}